\documentclass[fullpage]{article}
\usepackage{fullpage}
\usepackage{color}
\usepackage{authblk}
\usepackage{fancyhdr}
\usepackage{helvet}
\usepackage{amsmath}
\usepackage{psfrag}
\usepackage{setspace}
\usepackage[linktocpage]{hyperref}[2003/11/30]
\usepackage{lscape}
\usepackage{nicefrac}
\usepackage{mathrsfs}
\usepackage{units}
\usepackage{upgreek}
\usepackage{amssymb}
\usepackage{verbatim}
\usepackage[pdftex]{graphicx}
\usepackage{epstopdf}
\usepackage{hyperref}

\begin{document}

\title{Lorentz gauge quantization in synchronous coordinates}
\author{Christopher Garner and Dan N. Vollick}
\affil{Irving K. Barber School of Arts and Sciences, University of
British Columbia Okanagan\\
3333 University Way, Kelowna, British
Columbia V1V 1V7, Canada}
\maketitle

\begin{abstract}
\noindent It has been shown that the Gupta-Bleuler method of quantization can be used to impose the Lorentz gauge condition in static space-times but not in cosmological space-times. This implies that the Gupta-Bleuler approach fails in general in non-static space-times. More recently, however, the Dirac method of quantizing constrained dynamical systems has been successfully employed to impose the Lorentz gauge in conformally flat space-times. In this paper we generalize this result by using Dirac's method to impose the Lorentz gauge in a general space-time region where the metric is expressed in synchronous coordinates.
\end{abstract}

DOI: \href{http://dx.doi.org/10.1103/PhysRevD.93.064010}{\textcolor{blue}{http://dx.doi.org/10.1103/PhysRevD.93.064010}}

\newpage

\section{Introduction}

\noindent Dirac's approach to quantizing constrained dynamical systems [1,2] can be applied directly the problem of quantizing the source-free electromagnetic field. In this approach a gauge fixing term is added to the Lagrangian of the system. This breaks the gauge invariance of the Lagrangian as well as modifies the canonical momenta and the Hamiltonian. The gauge condition is directly enforced through constraints on the dynamic field variables.
\vspace{3mm}

\noindent To quantize the system the dynamic field variables are promoted to time-independent operators, while the Poisson brackets between dynamic field variables are replaced by commutators in the usual way. A wave function is introduced that evolves according to the Schrodinger equation using the modified Hamiltonian and is annihilated when acted on by the constraints.
\vspace{3mm}

\noindent The advantage of quantizing the electromagnetic field using Dirac's approach is that the Lorentz gauge condition can be imposed directly on the wave function. In the more popular Gupta-Bleuler formalism the Lorentz gauge condition is enforced more weakly by requiring only that the expectation value of the Lorentz gauge constraint vanish. It has been shown that one can use the Gupta-Bleuler approach to quantize the electromagnetic field in static space-times [3] but not in cosmological space-times [4]. This implies that the Gupta-Bleuler method fails in general in non-static space-times.
Recently the Dirac approach has been used to quantize the source-free electromagnetic field in general conformally flat space-times [5].

\vspace{3mm}
\noindent In this paper we apply the Dirac approach to a fully general synchronous space-time metric. This is of particular interest as any general metric can, at least locally, be re-expressed as a synchronous metric through a coordinate transformation [6]. Lastly, we define transverse and longitudinal components of the dynamic field variables as straight-forward covariant extensions of their Minkowski definitions. We show that only the transverse components of the dynamic field variables contribute weakly to the Hamiltonian and, under a certain assumption, to the energy-momentum tensor of the electromagnetic field. This shows that, in contrast to the Gupta-Bleuler approach, Dirac's method of quantization can be used in a general space-time region where the metric can be expressed in synchronous coordinates.

\section{Dirac Quantization in Synchronous Coordinates}
Consider a three dimensional space-like slice $\Sigma$ through the space-time without a boundary. Let the coordinates on the surface be denoted by $x^k$ (there may be more than one coordinate patch) and consider geodesics that pass through $\Sigma$ and are orthogonal to $\Sigma$. A point P in the neighborhood of $\Sigma$ is labeled by the proper time along the geodesic from the surface to P and by the coordinates $x^k$ at the point the geodesic intersects the $\Sigma$. This coordinate system will break down if geodesics that pass through different points on $\Sigma$ eventually cross. Therefore, these synchronous coordinates will, in general, only be defined in a neighborhood of $\Sigma$. In this neighborhood the metric takes the form
\begin{equation}
ds^2=-dt^2+h_{ij}(t,x)dx^idx^j.
\end{equation}

\noindent The Lagrangian will be taken to be the Lorentz gauge-fixed Lagrangian
\begin{equation}
\mathcal{L}=-\frac{1}{4}\sqrt{g}F_{\mu\nu}F^{\mu\nu}-\frac{1}{2}\sqrt{g}(\nabla_{\mu}A^{\mu})^2.
\label{eq:eq1}
\end{equation}
To obtain Maxwell's theory from this Lagrangian the constraint $\nabla_{\mu}A^{\mu}=0$ must be imposed.

\noindent The canonical momental density, $\frac{\partial\mathcal{L}}{\partial\dot{A}_{\mu}}$, is given by
\begin{equation}
 \Pi^{\mu}=-\sqrt{h}\left[F^{t\mu}+g^{t\mu}{\nabla}_{\nu}A^{\nu}\right].
\end{equation}

\noindent After integration by parts the Hamiltonian density, $\mathcal{H}= \Pi^{\mu}\dot{A}_{\mu}-\mathcal{L}$, that follows is
\begin{equation}
\mathcal{H}=\frac{1}{2}\frac{\Pi_{\mu}\Pi^{\mu}}{\sqrt{h}}+
\left[\,^{(3)}\nabla_kA^k-\frac{1}{2}h^{kl}\dot{h}_{kl}A_t\right]\Pi^t-A_t\partial_k\Pi^k+\frac{1}{4}\sqrt{h}F^{ij}F_{ij}
\end{equation}

\noindent where $^{(3)}\nabla_k$ is the covariant derivative on the three dimensional surface.

\noindent Our first constraint, given by the Lorentz condition, is
\begin{equation}
\chi_{1}=\frac{\Pi^t}{\sqrt{h}}\approx0.
\end{equation}

\noindent The symbol $\approx$ denotes a weak equality, requiring that the constraint be imposed \emph{after} any Poisson brackets have been evaluated.  For consistency we also require that
\begin{equation}
\dot{\chi}_{1}=\{\chi_{1},H\}+\frac{\partial\chi_1}{\partial{t}}\approx0
\end{equation}

\noindent where $H=\int{\mathcal{H}d^3x}$. For this consistency condition to be satisfied we require the additional secondary constraint
\begin{equation}
\chi_{2}=\frac{\partial_k\Pi^k}{\sqrt{h}}=\,^{(3)}\nabla_k\left(\frac{\Pi}{\sqrt{h}}^k\right)\approx0
\label{sec}
\end{equation}

 \noindent
 It is interesting to note that $\frac{\Pi^k}{\sqrt{h}}$ is a vector under the following transformation on the surface
\begin{equation}
\bar{t}=t\;\;\;\;\;\;\;\;\;\;\;\;\; \bar{x}^k=\bar{x}^k(x^l).
\label{trans}
\end{equation}
Applying the same consistency condition to $\chi_2$ we find that $\dot{\chi}_2\approx0$, thus there are no further constants. Both $\chi_1$ and $\chi_2$ are \emph{first-class} since $\{\chi_1,\chi_2\}=0$. The Hamiltonian density can be re-expressed in terms of our two constraints as
\begin{equation}
\mathcal{H}=\frac{1}{2}\frac{\Pi_{k}\Pi^{k}}{\sqrt{h}}+\frac{1}{4}\sqrt{h}F^{ij}F_{ij}+\Omega\chi_1-\sqrt{h}A_t\chi_2
\end{equation}

\noindent where $\Omega=\Pi_t/2+\sqrt{h}\left[\,^{(3)}\nabla_kA^k+\frac{1}{2}h^{kl}\dot{h}_{kl}A_t\right]$.

\noindent To quantize the the theory we promote the dynamic variables, $A_{\mu}$ and $\Pi^{\mu}$, to time-independent operators that satisfy
\begin{equation}
\left[A_{\mu}(\vec{x}),A_{\nu}(\vec{y})\right]=\left[\Pi^{\mu}(\vec{x}),\Pi^{\nu}(\vec{y})\right]=0
\end{equation}
and
\begin{equation}
\left[A_{\mu}(\vec{x}),\Pi^{\nu}(\vec{y})\right]=i\delta_{\mu}^{\nu}\delta(\vec{x},\vec{y}),
\end{equation}

\noindent where $[{\;}{}{\;}{\;}]$ denotes the commutator and we have set $\hbar=1$. A state vector is introduced which satisfies the Schrodinger equation
\begin{equation}
i\frac{d}{dt}\left|{\Psi}\right>=H\left|{\Psi}\right>.
\end{equation}

\noindent The constraints are then imposed on the wave function as follows:
\begin{equation}
\chi_1\left|{\Psi}\right>=0{\;}{\;}{\;}and{\;}{\;}{\;}\chi_2\left|{\Psi}\right>=0.
\end{equation}

\noindent We have chosen the operator ordering of the Hamiltonian such that the last two terms vanish upon application of our two constants and thus do not affect the equations of motion.

\section{Decomposition of the Hamiltonian into Transverse and Longitudinal Modes}

The Hamiltonian can be split as follows:
\begin{equation}
H=\int{\left[\frac{1}{2}\frac{\Pi_{k}\Pi^{k}}{\sqrt{h}}+\frac{1}{4}\sqrt{h}F^{ij}F_{ij}\right]d^3x}+\int{\left[\Omega\chi_1-\sqrt{h}A_t\chi_2\right]d^3x}.
\end{equation}

\noindent The second term weakly vanishes and can therefore be ignored since it vanishes when acting on physical states. We decompose the dynamic variables into transverse and longitudinal components
\begin{equation}
A_{k}=A_{k}^{(T)}+\,^{(3)}\nabla_kU{\;\;\;}{\;}{\;\;\;}and{\;\;\;}{\;}{\;\;\;}\Pi^{k}=\Pi^{k}_{(T)}+\,^{(3)}\nabla^k(\sqrt{h}V),
\label{dec}
\end{equation}

\noindent
where $U$ and $V$ are scalars that satisfy $^{(3)}\nabla^2U=\,^{(3)}\nabla^{k}A_k$ and $^{(3)}\nabla^2(\sqrt{h}V)=\,^{(3)}\nabla_{k}\Pi^k$ (we assume that solutions exist).
The transverse components then satisfy $^{(3)}{\nabla}^kA_{k}^{(T)}=$ $^{(3)}{\nabla}_k\Pi^{k}_{(T)}=0$. The non-weakly-vanishing term of the Hamiltonian can, after integration by parts, be decomposed into transverse and longitudinal components:
\begin{equation}
H_{(T)}=\int{\left[\frac{1}{2}\frac{\Pi^{(T)}_{k}\Pi_{(T)}^{k}}{\sqrt{h}}+\frac{1}{4}\sqrt{h}F^{ij}F_{ij}\right]d^3x}
\end{equation}
and
\begin{equation}
H_{(L)}=-\frac{1}{2}\int{V\,^{(3)}\nabla^2V\sqrt{h}d^3x},
\end{equation}

\noindent  where $^{(3)}\nabla^2V=\,^{(3)}\nabla_k\,^{(3)}\nabla^kV$ and $F_{ij}$ only contains the transverse components of the vector potential. In performing integration by parts we have dropped a surface term at infinity (if $\Sigma$ extends to infinity). This will be valid if it does not contribute to the classical equations of motion, which will be the case if the fields drop off sufficiently rapidly at infinity. From (\ref{sec}) and (\ref{dec}) it is easy to see that the constraint $\chi_2$ can be written as
\begin{equation}
\chi_2=\,^{(3)}\nabla^2V\approx0.
\end{equation}

\noindent We can therefore conclude that $H_{(L)}\approx0$ and that only the transverse components of dynamic variables contribute to the Hamiltonian.

\noindent Consider, in the classical theory, a surface of constant time, $t=t_0$, with $V_{t_{0}}(\vec{x})= V(\vec{x},t_{0})$ on the surface. Under certain conditions it is possible to prove that $^{(3)}\nabla^2V_{t_0}=0$ implies that $V_{t_0}=$ constant. Now this constant may vary from one surface to another implying that $V(\vec{x},t)=f(t)$, where $f(t)$ is a function of time not of $\vec{x}$. If $\Sigma$ extends to infinity $f(t)$ will vanish, since the fields vanish at infinity. Assuming that we can take the secondary constraint to be
$V(\vec{x},t)\approx f(t)$ we find that the longitudinal part of $\Pi^k$ weakly vanishes. This constraint can then be taken over to the quantum theory.

\section{The Energy-Momentum Tensor}

The energy-momentum tensor,
\begin{equation}
T^{\mu\nu}=\frac{2}{\sqrt{g}}\frac{\delta\mathcal{L}}{{\delta}g_{\mu\nu}}
\end{equation}

\noindent  satisfies
\begin{equation}
T^{\mu\nu}\approx T^{\mu\nu}_{Maxwell},
\end{equation}
where $T^{\mu\nu}_{Maxwell}$ is the energy-momentum tensor of Maxwell's theory. It can be shown that $T^{\mu\nu}$ contains only transverse
degrees of freedom, weakly, if $V(\vec{x},t)\approx f(t)$. Here we have ordered the operators so that the constraints appear on the right hand side of all expressions. It can also be shown that $H\approx-\int{T^t_{\:\:t}\sqrt{h}d^3x}$.

\section{Conclusion}

We have used Dirac's approach to quantizing constrained dynamical systems to quantize the source-free electromagnetic field in the Lorentz gauge in a general space-time expressed in synchronous coordinates. This generalizes the results of [5]. Consistency of the time evolution of the Lorentz constraint was ensured through the additional constraint $\partial_k\Pi^k/\sqrt{h}=0$. The two constraints are maintained under time evolution and impose the Lorentz condition in a general synchronous coordinate system. Any metric can be written in the form of Eq. (1) in a neighbourhood around a non-null hypersurface through the space-time [6]. This generalizes the quantization procedure to restricted regions of more general space-times.
We also found that only the transverse components of the dynamic field variables contribute weakly to the Hamiltonian and, assuming $V(\vec{x},t)\approx f(t)$, to the energy-momentum tensor of the source-free electromagnetic field.

\section*{Acknowledgements}
We would like to thank Don Witt for helpful comments on harmonic functions on Riemannian manifolds.
This research was supported by the Natural Sciences and Engineering Research Council of Canada.

\section*{References}

[1] P.A.M. Dirac, \textit{Lectures on Quantum Mechanics} (Dover Publications, 1964).
\vspace{2mm}

\noindent [2] P.A.M. Dirac, \textit{The Principles of Quantum Mechanics}, 4th edition (Oxford University press, 1958) Chapter 7.
\vspace{2mm}

\noindent [3] M.J. Pfenning, \textcolor{blue}{Phys. Rev. D \textbf{65}, 024009 (2001).}
\vspace {2mm}

\noindent [4] J. Beltran Jimenez, A.L. Maroto, \textcolor{blue}{Mod. Phys. Lett. A \textbf{26}, 3025 (2011); Prog. Theor. Phys. Suppl. \textbf{190}, 33 (2011);
AIP Conf. Proc. \textbf{1241}, 1033 (2010); Int. J. Mod. Phys. D. \textbf{18}, 2243 (2009); Phys. Lett. B \textbf{686}, 175 (2010).}

\vspace{1mm}

\noindent [5] D.N. Vollick, \textcolor{blue}{Phys. Rev. D \textbf{86}, 084057 (2012)};
J.C. Cresswell, D.N. Vollick, \textcolor{blue}{Phys. Rev. D \textbf{91}, 084008 (2015)}.
\vspace{2mm}

\noindent [6] R.M. Wald, \textit{General Relativity}, (University of Chicago press, 1984) Section 3.3.

\end{document}